\documentclass[letterpaper, 10 pt, conference]{ieeeconf}  

\IEEEoverridecommandlockouts                              

\usepackage{graphics} 
\usepackage{epsfig} 
\usepackage{mathptmx} 
\usepackage{times} 
\usepackage{amsmath} 
\usepackage{amssymb}  
\usepackage{booktabs}
\usepackage{footnote}
\usepackage{threeparttable}

\title{\LARGE \bf
3D Gaussian Splatting for Fine-Detailed Surface Reconstruction in Large-Scale Scene
}

\author{Shihan Chen$^{1}$, Zhaojin Li$^{1}$, Zeyu Chen$^{1}$, Qingsong Yan$^{2}$, Gaoyang Shen$^{2}$, and Ran Duan$^{1, *}$
\thanks{$^{1}$ S. Chen, Z. Li, Z. Chen and R. Duan are with the Department of Land Surveying and Geo-Informatics, The Hong Kong Polytechnic University, Hong Kong, China. Email: {\tt\small \{shihan.chen, zhaojin.li, zeyu-andrew.chen\}@connect.polyu.hk, rduan@polyu.edu.hk}.}
\thanks{$^{2}$ Q. Yan and G. Shen are with the School of Geodesy and Geomatics, Wuhan University, Wuhan, China. Email: {\tt\small yanqs.whuer@gmail.com, lamb31832@gmail.com}.}
\thanks{$^{*}$ indicates the corresponding author.}}

\begin{document}

\maketitle
\thispagestyle{empty}
\pagestyle{empty}

\begin{abstract}
\label{sec:abstract}
Recent developments in 3D Gaussian Splatting have made significant advances in surface reconstruction. However, scaling these methods to large-scale scenes remains challenging due to high computational demands and the complex dynamic appearances typical of outdoor environments. These challenges hinder the application in aerial surveying and autonomous driving. This paper proposes a novel solution to reconstruct large-scale surfaces with fine details, supervised by full-sized images. Firstly, we introduce a coarse-to-fine strategy to reconstruct a coarse model efficiently, followed by adaptive scene partitioning and sub-scene refining from image segments. Additionally, we integrate a decoupling appearance model to capture global appearance variations and a transient mask model to mitigate interference from moving objects. Finally, we expand the multi-view constraint and introduce a single-view regularization for texture-less areas. Our experiments were conducted on the publicly available dataset GauU-Scene V2, which was captured using unmanned aerial vehicles. To the best of our knowledge, our method outperforms existing NeRF-based and Gaussian-based methods, achieving high-fidelity visual results and accurate surface from full-size image optimization. Open-source code will be available on GitHub.

\end{abstract}


\section{INTRODUCTION}
\label{sec:intro}


3D reconstruction in large-scale scenes enables several essential applications in domains like 3D mapping and autonomous driving \cite{Vizzo2021Poisson, Liu2019HighDefinitionMap}. For example, a high-fidelity map can serve as a simulation environment for self-driving vehicles, and a highly accurate geometry map can be a prior for navigation \cite{tancik2022block, Jun2024RenderableStreetView, Yan2024StreetGaussians}. The photo-realistic rendering is essential for simulation environments. Although traditional methods can efficiently reconstruct large-scale surfaces, they still fall short in visual quality and may result in discontinuous or overly smoothed surfaces \cite{chen2019learning}.

Recently, Neural Radiance Field (NeRF) \cite{mildenhall2021nerf} and subsequent methods \cite{barron2021mip, muller2022instant, turki2022mega} have been applied to various tasks, including accurate reconstruction of surfaces \cite{NEURIPS2021_e41e164f, fu2022geo, li2023neuralangelo}. However, the requirement of considerable time and computational resources limits their application, making them struggle to achieve large-scale reconstruction and real-time rendering. The emergence of 3D Gaussian Splatting (3DGS) \cite{kerbl20233d} offers a promising solution to these challenges. 3D Gaussian is a flexible and impressive explicit representation, significantly decreasing computational requirements \cite{lin2024gaussian, yu2024mip}. Based on 3DGS, several remarkable surface reconstruction methods have been developed \cite{guedon2024sugar, huang20242d, chen2024pgsr, yu2024gaussian}. However, these methods primarily focus on small-scale and object-centric scenes. Applying them directly to large-scale scenes may result in convergence difficulties, out-of-memory errors, and additional Gaussian primitives due to appearance variations and transient objects in outdoor environments \cite{lin2024vastgaussian}. Some variants of 3DGS have achieved excellent real-time rendering results in large-scale scenes \cite{lin2024vastgaussian, liu2024citygaussian, chen2024dogaussian}. They face challenges in reconstructing accurate geometric surfaces, and the images used for optimization need to be downsampled, resulting in significant information loss. Additionally, it is easy to result in local optima by optimizing Gaussian primitives solely supervised by the color of images, ultimately failing to restore actual surfaces due to the disorderly and irregular nature of 3D Gaussians \cite{huang20242d}.

\begin{figure}[t] 
	\centering
	\includegraphics[width=\linewidth,scale=1.00]{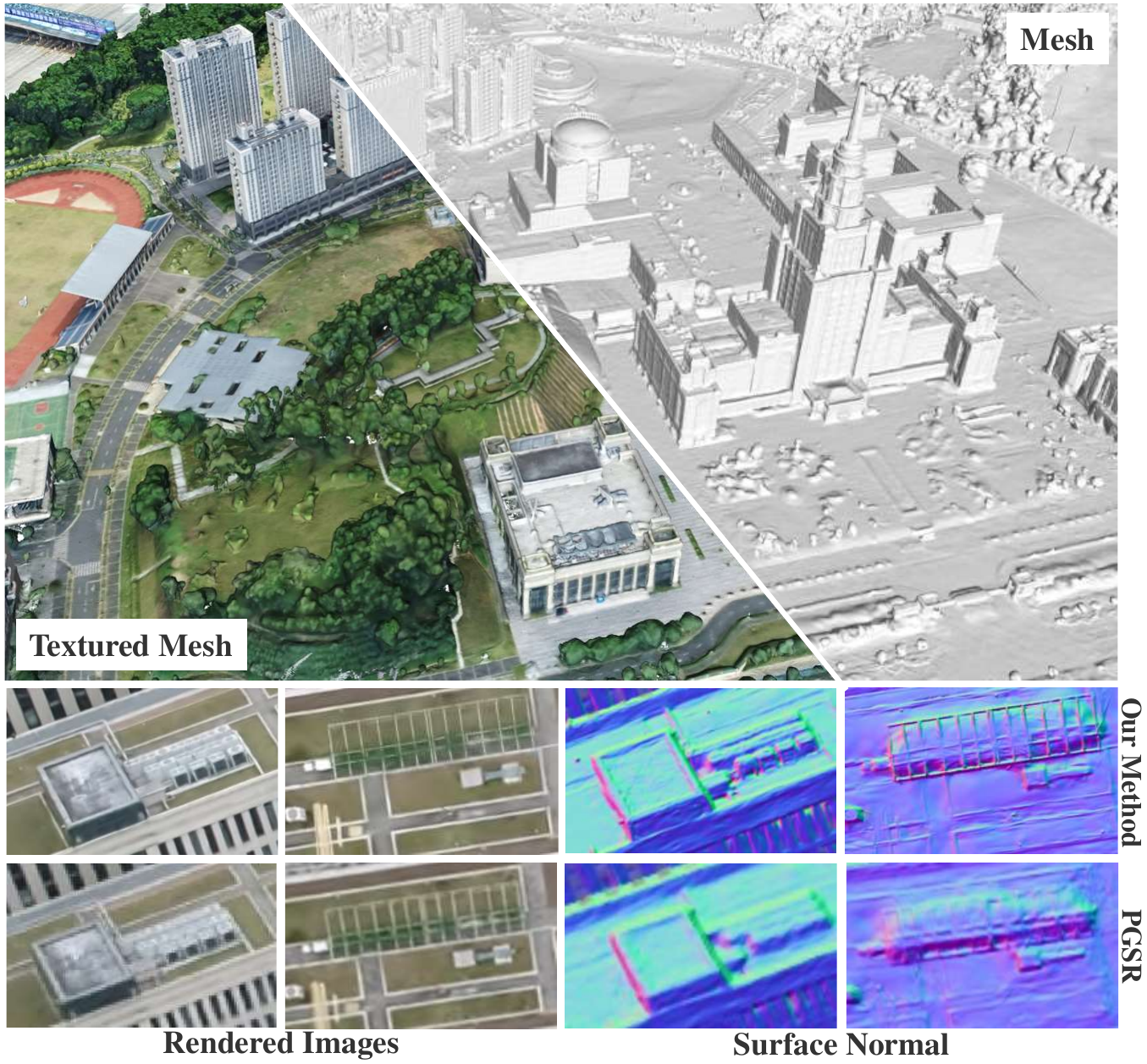}
        \vspace{-22 pt}
	\caption{ Our method can reconstruct large-scale surface models and outperforms the SOTA Gaussian-based surface reconstruction method PGSR \cite{chen2024pgsr} with higher visual quality and level of detail.}
        \vspace{-23 pt}
	\label{Figure0}
\end{figure}

To enhance Gaussian-based approaches for meshing high-fidelity surfaces in large-scale scenes, we propose a framework based on planar-based Gaussian splatting \cite{chen2024pgsr}. Firstly, for numerous computational resources by processing large amounts of full-size images, we propose a coarse-to-fine strategy to reconstruct a coarse Gaussian model from downsampled images followed by adaptive scene partitioning and refining by split images in parallel. Unlike regular division, which leads to an uneven distribution of images between sub-scenes, our approach partitions the scene by a binary tree structure, selects images based on visibility, and determines whether further partitioning is necessary according to the selected image number and cell size. In the refining stage, we split the images and utilize one sub-image in one optimization iteration to avoid out-of-memory errors.

Secondly, we introduce a decoupling global appearance model to fit significant illumination variations and a transient mask model to remove moving objects from images. Following VastGaussian \cite{lin2024vastgaussian}, we feed the rendered image into a Convolutional Neural Network (CNN) to learn the global appearance transformation compared to ground truth (GT) images, allowing the Gaussian model to focus on the geometric reconstruction and the static appearance. As for transient objects, we employ a Unet model \cite{ronneberger2015u} to generate a 2D visibility map for captured images, facilitating accurate segmentation between transient and static objects and only optimizing static objects.

Furthermore, multi-view geometric and photometric constraints have been evaluated to obtain consistent and robust results \cite{fu2022geo, schonberger2016pixelwise}, which are also introduced in our method. However, due to a lack of constraints in texture-less areas, like water bodies and some roofs, the Gaussian model tends to grow uncontrollably. We propose a plane regularization by reducing the depth gradients in texture-less areas of images. To sum up, our contributions are listed as follows:

\begin{itemize}
\item To reconstruct large-scale scenes, we propose a coarse-to-fine strategy with adaptive scene partitioning.
\item To obtain a consistent appearance and geometry, we integrate a decoupling appearance model to fit global uneven illumination and a transient mask model for dynamic object removal.
\item To maintain accurate geometry with fine details, we enhance transient-aware multi-view constraints, introduce a textureless area regularization, and utilize full-resolution images in the refinement stage.
\end{itemize}
\section{Related Work}
\label{sec:Related work}
\begin{figure*}[htbp] 
	\centering
        \vspace{3 pt}
	\includegraphics[width=0.75\linewidth,scale=1.00]{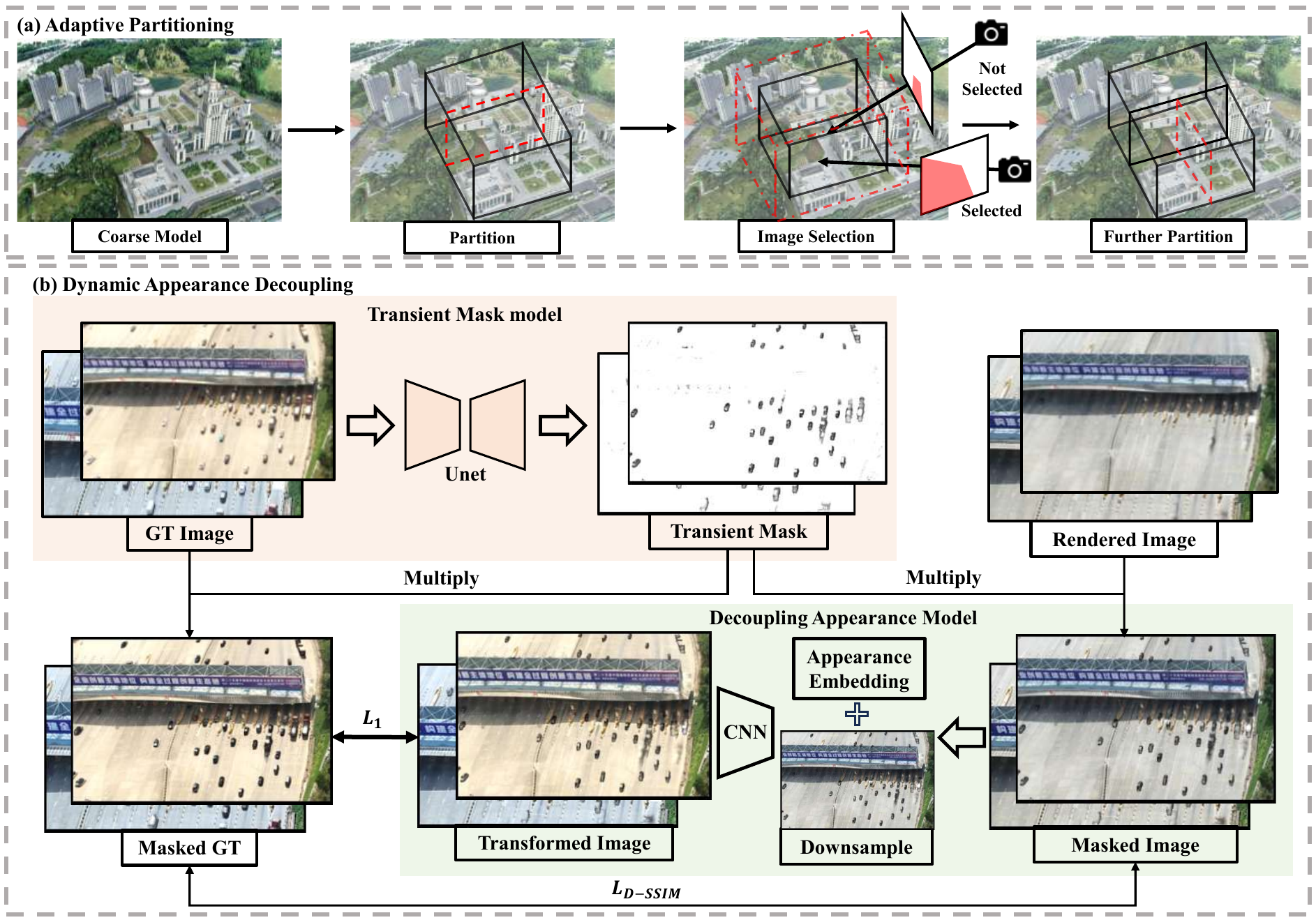}
        \vspace{-10 pt}
	\caption{ Illustration of adaptive partitioning and dynamic appearance decoupling. (a) We first reconstruct coarse 3D Gaussians from downsampled images, adaptively partition the Gaussian model according to image number and cell size, and refine sub-scenes using full-size images. (b) For Gaussian reconstruction, an Unet model facilitates the segmentation of transient and static objects, and transient objects are removed in optimization. A CNN network transforms the rendered image from 3D Gaussians to learn global appearance variations.}
        \vspace{-15 pt}
	\label{Figure1}
\end{figure*}
\subsection{Novel View Synthesis}
The emergence of NeRF \cite{mildenhall2021nerf, barron2021mip, muller2022instant, turki2022mega, NEURIPS2021_e41e164f} has attracted the attention of academia and industry due to its ability to synthesize photo-realistic novel views. However, the numerous training and rendering time limits the application. To decrease the computational consumption, Plenoxels \cite{fridovich2022plenoxels} and Instant-ngp \cite{muller2022instant} introduce gird-based representation instead of fully MLPs. 3DGS \cite{kerbl20233d, huang20242d, chen2024pgsr, lin2024vastgaussian, wu20244d} is another promising approach in novel view rendering by optimizing a set of 3D Gaussian primitives. Without neural networks, it and the subsequent works can achieve fast optimization and real-time rendering. Mip-Splatting \cite{yu2024mip} addresses the aliasing problems by introducing low-pass filters to constrain the Gaussian scale. LightGaussian \cite{fan2023lightgaussian} compacts 3D Gaussian to reduce storage usages, and 4D Gaussian \cite{wu20244d} promotes Gaussian splatting to dynamic scenes.

\subsection{Surface Reconstruction}
NeRF-based methods are combined with signed distance function (SDF) to reconstruct high-fidelity surfaces due to the continuous and differentiable nature \cite{NEURIPS2021_e41e164f, fu2022geo, li2023neuralangelo, oechsle2021unisurf}. NeuS \cite{NEURIPS2021_e41e164f} promotes an unbiased mapping between SDF and volume rendering, resulting in an accurate surface reconstruction. Neuralangelo \cite{li2023neuralangelo} replaces analytical gradients with numerical gradients, guiding SDF to reconstruct higher levels of detail from coarse to fine. As for Gaussian splatting, notable works align Gaussian primitives with actual surfaces and extract accurate meshes from them \cite{guedon2024sugar, huang20242d, chen2024pgsr, yu2024gaussian}. SuGaR \cite{guedon2024sugar} is the first work to enforce the alignment of 3D Gaussians with the surface, and 2DGS \cite{huang20242d} collapses one dimension of the 3D Gaussian. PGSR \cite{chen2024pgsr} further applies the multi-view geometric and photometric constraints in optimization, improving surface accuracy. While these solutions mainly concentrate on the object-level scenes, applying them to large-scale surface reconstruction remains a limitation.

\subsection{Large-scale Reconstruction}
Some subsequent works of NeRF \cite{tancik2022block, turki2022mega, zhenxing2022switch, xu2023grid, xiangli2022bungeenerf} and 3D Gaussian \cite{lin2024vastgaussian, chen2024dogaussian, liu2024citygaussian} are scaled to large-scale scene. Block-NeRF \cite{tancik2022block} and Mega-NeRF \cite{turki2022mega} divide a large-scale scene and train each cell in parallel. Grid-NeRF \cite{xu2023grid} introduces a feature grid representation rather than scene partitioning with sub-NeRF for each cell. VastGaussian \cite{lin2024vastgaussian} proposes a data partitioning strategy and further addresses the variant appearance, refining the visual quality of large-scale scenes. CityGaussian \cite{liu2024citygaussian} combines a level-of-detail strategy with data partitioning, enabling efficient rendering. CityGaussianV2 \cite{liu2024citygaussianv2} scales 2DGS to reconstruct large-scale surfaces but still faces challenges with full-resolution images. Our work reconstructs large-scale surfaces from full-sized images while also considering appearance variations and texture-less areas.
\section{Preliminaries}
3DGS \cite{kerbl20233d} represents a scene as Gaussian primitives, defined by a position, covariance matrix, opacity, and spherical harmonics (SH) coefficients for view-dependent radiance. In rendering, images are generated by projecting and $\alpha$-blending Gaussian primitives as shown in Eq. (\ref{eq:a-blending}):
\begin{equation}
  C=\sum_{i \in W} \alpha_{i} c_{i} \prod_{j=1}^{i-1}\left(1-\alpha_{j}\right),
  \label{eq:a-blending}
\end{equation}
where $C$ represents the rendered color, $c_{i}$ is the color of the $i$-th Gaussian primitive determined by SH coefficients and view direction, and $\alpha_{i}$ is derived from the opacity and covariance matrix.

Compared to 3D Gaussians, with difficulty adhering to actual surfaces, we utilize planar-based Gaussian splatting\cite{chen2024pgsr} and employ unbiased depth rendering to obtain depth and surface normal maps. PGSR flattens the Gaussian primitive along the direction of minimum scale by $L_{flatten}$. This direction is also defined as the normal of the Gaussian plane denoted as $n_{i}$, followed by $\alpha$-blending to achieve the view-dependent surface normal map:
\begin{equation}
  N=\sum_{i \in W} R_{c}^{w} n_{i} \alpha_{i} \prod_{j=1}^{i-1}\left(1-\alpha_{j}\right),
  \label{eq:n-blending}
\end{equation}
where $R_{c}^{w}$ denotes the rotation matrix from camera to world. PGSR assumes that during the optimization process, the Gaussian plane will gradually fit onto the actual surface, and it employs $\alpha$-blending to acquire the distance from the surface patch to the camera center. Subsequently, the depth map could be determined by intersecting rays with the surface patch, denoted as $D$:
\begin{equation}
  D=\frac{ \sum_{i \in W} d_{i} \alpha_{i} \prod_{j=1}^{i-1}\left(1-\alpha_{j}\right)}{N(p) K^{-1} \tilde{p}},
  \label{eq:depth}
\end{equation}
where $d_{i}$ is the distance between camera center and $i$-th Gaussian plan, $p$ represents the 2D coordinate in image plane, $\tilde{p}$ is the corresponding homogeneous coordinate, and $K$ is the intrinsic matrix of the camera.
\section{Method}
This paper proposes a divide-and-conquer manner with a Gaussian model refining from full-size images and decoupling the complex dynamic appearance. In addition to image constraints, constraints directed to geometry are also introduced to ensure geometric consistency.

In \ref{Adaptive Partitioning}, we introduce an adaptive data partitioning method based on each cell's image number and size. In Sec. \ref{Dynamic Appearance}, the global appearance model captures the variant appearance in images, and the transient removing model excludes moving objects from optimization. Sec. \ref{Geometry Regularization} describes multi-view constraints and texture-less area regularization. Finally, we detail mesh extraction and Gaussians merging in Sec. \ref{Mesh Extraction}.

\subsection{Adaptive Partitioning} \label{Adaptive Partitioning}

\begin{figure*}[ht] 
	\centering
	\includegraphics[width=\linewidth,scale=1.00]{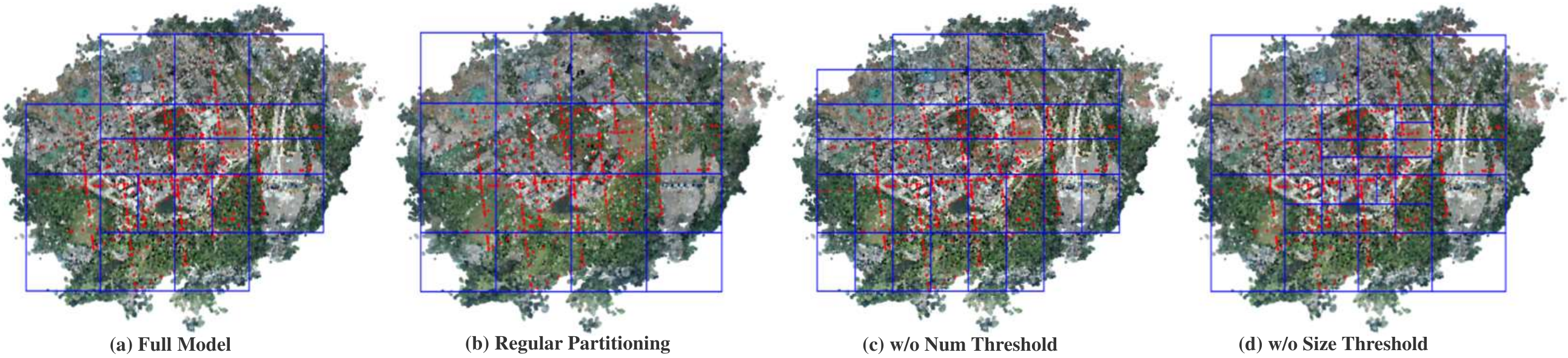}
        \vspace{-20 pt}
	\caption{Visualization of different partitioning strategies. The red points are the camera positions, and the blue boxes are the cell edges.}
        \vspace{-18 pt}
	\label{Figure_3}
\end{figure*}

Directly optimizing a large-scale scene from full-size images leads to heavy computational consumption and difficulties in convergence. We utilize a coarse-to-fine strategy to reconstruct the coarse Gaussian model, partition it, and refine every cell from split images. With a coarse model and fewer images, our algorithm can further restore the fine details with lower computational resources. Fig. \ref{Figure_3}(b) indicates that naively using regular partitioning may result in some cells containing very few images, and we propose an adaptive partitioning method. The overview is shown in Fig. \ref{Figure1}(a).

We divide the high-resolution image into sub-images to facilitate accurate scene partitioning. For large-scale scenes, we assume that the ground plane has been orientated parallel to the $xy$ plane of the coordinate. Our method projects the Gaussian primitives on the ground plane and divides the scene into two parts at the midpoint of the long axis. The bounding box length of $i$-th cell is denoted as $(l_{i,x}, l_{i,y})$. To prevent color discrepancies and geometric misalignments in merging, we expand boundaries by a certain percentage. Then, all the images $I_{i}$ and Gaussian primitives $G_{i}$ located in this range are selected, and the 3D bounding box is determined. Reconstruction with images selected by position may result in floaters or poor details due to insufficient supervision. We project the $i$-th bounding box to the $j$-th image outside the cell, with the projected area $S_{i,j}$. If the ratio of $S_{i,j}$ to the entire area of the image exceeds a certain threshold, the image is added to $I_{i}$.

Employing partitioning only once may not fulfill the computational requirement. Those cells with image numbers greater than the threshold $N$ and the minimum length of $l_{x}$ and $l_{y}$ exceeding the threshold $L$ are further partitioned. Partitioning based solely on cell size may result in some sub-cells containing few images, as demonstrated in Fig. \ref{Figure_3}(c). Partitioning based solely on the number of images can lead to excessive partitioning in areas with dense image coverage, as shown in Fig. \ref{Figure_3}(d). On the other hand, some sub-cells at the boundaries of a scene may still lack a sufficient number of images, which will be removed, as shown in Fig. \ref{Figure_3}(a). Our algorithm repeats this process until each cell meets the required image quantity or range criteria. Finally, every Gaussian primitive visible to images $I_{i}$ is added to set $G_{i}$. 

Subsequently, our method refines each model with the supervision of split images without downsampling, facilitating pixel-wise surface reconstruction. We calculate the perspective projection matrix from intrinsics but not the field of view to support noncentral principal point image rendering.

\subsection{Dynamic Appearance Decoupling} \label{Dynamic Appearance}

Outdoor scene images are usually captured with uneven environmental illumination and exposure, leading to an apparent dynamic appearance, including global variation and transient appearance. The global variant appearance can lead to low opacity 3D Gaussians near the camera \cite{lin2024vastgaussian}, and gradients near the transient object regions can result in meaningless growth of Gaussian primitives \cite{Zhang2024Gaussianinthewild}. We introduce an appearance model to fit the global variation and a transient mask model to remove moving vehicles and pedestrians.

\paragraph{Decoupling Appearance Model} To ensure 3D Gaussians focus on geometric information, we decouple the appearance by adopting a CNN \cite{lin2024vastgaussian}. As shown in Fig. \ref{Figure1}(b), the rendered images $I_{(i,r)}$ from 3D Gaussians are downsampled and concatenated with an appearance embedding $l_{i}$. The 2D map is fed into a CNN and progressively upsampled to $T_{i}$ with the same resolution as $I_{(i,r)}$. Then, our method performs a pixel-wise multiplication on $T_{i}$ and generates an appearance-transformed image $I_{(i, a)}$, which is compared with GT image $I_{i}$ to calculate the ${L}_{1}$ loss. To ensure the structure similarity, The ${L}_{D-SSIM}$ is applied between $I_{(i,r)}$ and $I_{i}$ as shown in Eq. \ref{eq:L_rgb_mask}.
The CNN and appearance embeddings are both optimized in reconstruction. By first downsampling to remove most high-frequency information, the CNN can concentrate on learning global patterns, facilitating the Gaussian model to render appearance-consistent images.

\paragraph{Transient Mask Model} To avoid the emergence of additional Gaussian primitives introduced by transient objects, our method segments transient objects and static objects by employing a Unet model to generate a 2D mask $M_{i}$ \cite{ronneberger2015u, Zhang2024Gaussianinthewild}. The Unet model takes a GT image $I_{i}$ as input and produces a same-size mask with one as static objects and zero as transient objects. In calculating the ${L}_{1}$ and ${L}_{D-SSIM}$, the mask is multiped on $I_{i}$, $I_{(i,r)}$ and $I_{(i,a)}$ to reduce the large losses in areas containing transient objects denoted in Eq. \ref{eq:L_rgb_mask}. To prevent marking all pixels as transient, we introduce a regularization term as shown in Eq. \ref{eq:L_mask}. In this way, an adversarial framework between ${L}_{rgb}$ and ${L}_{mask}$ unsupervised removes objects captured a few times at a position.
\begin{equation}
\begin{aligned}
{L}_{rgb}=(1-\lambda_{1}) {L}_{1}\left(M_{i} \cdot I_{(i,a)}, M_{i} \cdot I_{i}\right)+ \\ \lambda_{1} {L}_{D-SSIM}\left(M_{i} \cdot I_{(i,r)}, M_{i} \cdot I_{i}\right),
  \label{eq:L_rgb_mask}
\end{aligned}
\end{equation}
\begin{equation}
{L}_{mask}=\lambda_{2} {L}_{1}\left(M_{i}, 0\right),
  \label{eq:L_mask}
\end{equation}
where $\lambda_{1}$ and $\lambda_{2}$ are weights in final training loss, 0.25 and 0.8 in our experiments, respectively.

\subsection{Geometry Regularization} \label{Geometry Regularization}

\begin{table*}[ht]
\centering
\vspace{4 pt}
\caption{Quantitative comparison of geometric accuracy. The MAE$\downarrow$ and RMSE$\downarrow$ are magnified 100 times.}
\vspace{-5 pt}
\setlength{\tabcolsep}{0.25cm}
\vspace{-5 pt}
\scalebox{0.9}{\begin{tabular}{l|cc|cc|cc|cc|cc|cc}
\toprule[1pt]

 Scene & \multicolumn{2}{c|}{Neuralangelo \cite{li2023neuralangelo}} & \multicolumn{2}{c|}{2DGS \cite{huang20242d}} & \multicolumn{2}{c|}{GOF \cite{yu2024gaussian}}  & \multicolumn{2}{c|}{PGSR \cite{chen2024pgsr}} & \multicolumn{2}{c|}{VastGaussian \cite{lin2024vastgaussian}} & \multicolumn{2}{c}{Our method} \\ \midrule
 
 Metrics & MAE & RMSE & MAE & RMSE & MAE & RMSE & MAE & RMSE & MAE & RMSE & MAE & RMSE     \\ \midrule
Campus             & 3.55 & 5.17 & 2.77 &  \textbf{2.56} & \underline{2.55} & 3.36     & \textbf{2.44} &3.20    & 3.03 & 4.19 &  \textbf{2.44}  & \underline{2.94} \\

Modern Building    & 2.86 & 3.46 & 27.13 & 33.90    & 9.66  & 15.34   & \underline{1.89} & \underline{3.17} & 6.97 & 6.80 & \textbf{1.82}  & \textbf{2.67} \\

Village            & 2.82 & 4.45 & 1.34 & 3.60     & \textbf{1.11} & \textbf{1.80}  &  \underline{1.15}    &3.98 & 1.63 & 2.27 & \underline{1.15} & \underline{2.22}  \\

Residence          & 4.23 & 4.48 & 1.99 & 6.12 &  \underline{1.50}  &  \textbf{2.38} &1.67 &2.87 & 1.84 & 2.79 & \textbf{1.46} & \underline{2.71}  \\

Russian Building    & 4.84 & 3.63 & 3.99 & 3.56 & 4.06 & 3.25 &  \underline{0.69} &  \underline{1.55}  & 2.23 & 3.13 & \textbf{0.67} & \textbf{1.43} \\

College             & 3.71 & 4.15 & \textbf{0.89} & \textbf{1.04}  & 1.34 & 1.99 &  \underline{1.04}  &  \underline{1.55} & 6.05 & 7.17 & \underline{1.04} &1.58 \\

Technology College  & 2.39 & 2.66 & 1.40          & 3.41     & 1.12          & 1.67     & \underline{0.84} & \underline{1.54} & 1.62 & 1.96 & \textbf{0.78}  & \textbf{1.38} \\

\midrule
Mean & 3.49 & 4.00 & 5.64 & 7.74 & 3.05 & 4.26 & \underline{1.42} & \underline{2.55} & 3.25 & 3.90 & \textbf{1.34}  & \textbf{2.13} \\
\bottomrule[1pt]
\end{tabular} }
\label{Table_1}
\end{table*}

\begin{table*}[ht]
\centering
\caption{Quantitative comparison of visual quality. We report SSIM$\uparrow$ , PSNR$\uparrow$ and LPIPS$\downarrow$.}
\vspace{-10 pt}
\setlength{\tabcolsep}{3pt}
\scalebox{0.9}{\begin{tabular}{l|ccc|ccc|ccc|ccc|ccc|ccc}
\toprule[1pt]
 Scene            & \multicolumn{3}{c|}{Neuralangelo \cite{li2023neuralangelo}} & \multicolumn{3}{c|}{2DGS \cite{huang20242d}} & \multicolumn{3}{c|}{GOF \cite{yu2024gaussian}}               & \multicolumn{3}{c|}{PGSR \cite{chen2024pgsr}} & \multicolumn{3}{c|}{VastGaussian \cite{lin2024vastgaussian}} & \multicolumn{3}{c}{Our method} \\ \midrule
Metrics & SSIM   & PSNR   & LPIPS & SSIM   & PSNR   & LPIPS  & SSIM          & PSNR  & LPIPS         & SSIM   & PSNR   & LPIPS  & SSIM   & PSNR   & LPIPS & SSIM          & PSNR           & LPIPS         \\ \midrule
Campus             & 0.40   & 18.91 & 0.70 & 0.58   & 23.32  & 0.54   &  \textbf{0.63} & 21.79 &  \textbf{0.49} & 0.58   & \underline{23.84} & 0.52 & 0.55 & 23.61 & 0.53 & \underline{0.61} & \textbf{24.67} & \underline{0.50}\\
Modern Building    & 0.42   & 20.90  & 0.65 & 0.50   & 18.31  & 0.62   & 0.54 & 20.79 & 0.60 &  \underline{0.62}   & \underline{23.77} & \underline{0.49} & 0.59 & 23.02 & \underline{0.49} & \textbf{0.66} & \textbf{25.04} & \textbf{0.46} \\
Village             & 0.46   & 20.91  & 0.62 & 0.62   & 23.43  & 0.51   & \underline{0.67} & \underline{25.31} & \underline{0.45} & 0.64   & 24.13  & 0.47 & 0.64 & 23.01 & \underline{0.45}  & \textbf{0.70} & \textbf{26.41} & \textbf{0.43} \\
Residence          & 0.32 & 17.28 & 0.68 & 0.54   & 21.62  & 0.57   & 0.57  & 22.51 & 0.53 & 0.57 & \underline{22.68} & 0.52 & \underline{0.58} & 21.99 & \underline{0.50} & \textbf{0.61} & \textbf{23.92} & \textbf{0.49} \\
Russian Building   & 0.45 & 18.06 & 0.61 & 0.62 & \underline{23.42} & 0.50  & 0.62  & 21.16 & 0.49 & 0.61   & 22.25  & 0.47 & \underline{0.64} & 23.07 & \underline{0.45} & \textbf{0.68} & \textbf{25.79} & \textbf{0.41} \\
College            & 0.41   & 19.17 & 0.60 & 0.64   & 22.79  & 0.51   & \underline{0.65} & 23.53 & 0.50 & \underline{0.65} & \underline{23.69} & \underline{0.49} & 0.56 & 21.64 & \underline{0.49} & \textbf{0.69} & \textbf{24.94} & \textbf{0.46} \\

Technology College & 0.44 & 18.56 & 0.62 & 0.59   & 22.27  & 0.55   & 0.68  & \underline{22.38} & 0.45 & 0.69 & 23.23  & 0.44 & \underline{0.71} & 23.39 & \underline{0.40}  & \textbf{0.75} & \textbf{25.02} & \textbf{0.39} \\ \midrule

Mean & 0.41 & 19.11  & 0.64 & 0.58   & 22.16  & 0.54   & \underline{0.64}  & 22.87 &  0.48 & 0.62 &  \underline{23.37}  & 0.49 & 0.61 & 22.82 & \underline{0.47} & \textbf{0.67} & \textbf{25.11} & \textbf{0.45} \\
\bottomrule[1pt]
\end{tabular}}
\vspace{-15 pt}
\label{Table_2}
\end{table*}

\paragraph{Multi-view Constraint} Due to the discrete and disordered nature of Gaussians, relying solely on image-based supervision is insufficient. We adopt multi-view geometric and photometric constraints to optimize geometry directly. Our method obtains the surface patch described by depth $D$ and normal $N$. Plane-induced homography \cite{schonberger2016pixelwise} defines the mapping between pixel patch from source image $I_{s}$ and a corresponding patch from reference image $I_{r}$:
\begin{equation}
\begin{aligned}
H_{sr} =  K_{r} & (R_{sr}-\frac{T_{sr}N_{s}^{T}}{D_{s}})K_{s}^{-1},
\label{eq:homography_matrix}
\end{aligned}
\end{equation}
\begin{equation}
\begin{aligned}
\tilde{p_{r}} & = H_{sr} \tilde{p_{s}},
\end{aligned}
  \label{eq:homography}
\end{equation}
where $R_{sr}$ and $T_{sr}$ are the relative transformation of rotation and translation from $I_{r}$ to $I_{s}$, $K$ is intrinsic matirx, and $\tilde{p_{r}}$ and $\tilde{p_{s}}$ are the homogeneous coordinate of pixel patches. It is similar to finding the coordinate $\tilde{p_{r,s}}$ by reprojecting $\tilde{p_{r}}$ to the source image utilizing $H_{rs}$. Due to the inaccurate depth and normal map in optimizing, the distance between $\tilde{p_{r,s}}$ and $\tilde{p_{s}}$ can be calculated as multi-view geometry loss:
\begin{equation}
{L}_{mvgeo} = \frac{1}{|L|} \sum_{\tilde{p_{s}} \in L} {L}_{1}(\tilde{p_{s}}, H_{rs} H_{sr} \tilde{p_{s}}),
  \label{eq:L_geometry}
\end{equation}
where $L$ is the valid overlapping region between source and reference images. The reprojected error exceeding a certain threshold may be caused by occlusion or structural geometric errors filtered in calculating.

In addition, the color of the same surface tends to be similar when projected to two different views, as do the source and reference pixel patches. Therefore, our method measures their similarity by Zero Mean Normalized Cross-Correlation (ZNCC) \cite{yoo2009fast}. To deal with occlusions and improve the robustness, we utilize a transient mask to eliminate interference from moving objects in calculating photometric consistency loss:
\begin{equation}
{L}_{zncc} = \frac{1}{|L|} \sum_{\tilde{p_{s}} \in L} {L}_{1} (1, ZNCC(M_{s} \tilde{p_{s}}, M_{r} \tilde{p_{r}})).
\vspace{-5 pt}
  \label{eq:L_photometric}
\end{equation}

\paragraph{Texture-less Area Constraint} Although multi-view constraints can ensure geometric consistency, they can hardly handle the uncontrolled cloning and splitting of 3D Gaussians in texture-less areas. The reprojected error may exceed the threshold, and the photometric constraint could capture a few textures. The texture-less regions, including water bodies and roofs, are usually smooth without sudden depth changes. Therefore, we propose a simple and effective way to minimize the depth gradient in that region. This constraint is invalid for edges and sudden depth discontinuities between objects and backgrounds. As image edges are usually related to geometric edges, we adopt image gradient $\nabla I$ to weight texture-less area constraint:
\begin{equation}
{L}_{gran} = \frac{1}{|I|} \sum_{i \in I} |1 - \nabla I(i)|^{5} | \nabla D(I).
\vspace{-5 pt}
  \label{eq:L_textureless}
\end{equation}

\paragraph{Depth-normal Regularization} Following PGSR  \cite{chen2024pgsr}, we supervise the rendered normal map by normal calculated from the unbiased depth. For a pixel $p_{i}$ in the depth map, our approach projects four neighboring pixels to the camera coordinate as 3D points $P_{i,0 \cdots 3}$. The normal derived from the depth map, as defined in Equation Eq. \ref{eq:Depth_normal}, is utilized to constrain the normal map. Additionally, image gradient is employed to weight it as shown in Eq. \ref{eq:L_consistency}:
\begin{equation}
N(d)_{i} = \frac{(P_{i,0}-P_{i,1}) \times (P_{i,3}-P_{i,2})}{|(P_{i,0}-P_{i,1}) \times (P_{i,3}-P_{i,2})|} ,
  \label{eq:Depth_normal}
\end{equation}
\begin{equation}
{L}_{cons} = \frac{1}{|I|} \sum_{i \in I} |1 - \nabla I(i)|^{5} {L}_{1}( N(d)_{i},N_{i} ).
  \label{eq:L_consistency}
\end{equation}

Our final constraints on geometry ${L}_{geo}$ consists of multi-view constraint, depth gradient constraint, and depth-normal regularization, as demonstrated in Eq. \ref{eq:L_geo}, and the final training loss is denoted in Eq. \ref{eq:L}:
\begin{equation}
{L}_{geo} = \lambda_{3} {L}_{cons} + \lambda_{4}{L}_{mvgeo} + \lambda_{5}{L}_{zncc} + \lambda_{6}{L}_{gran},
  \label{eq:L_geo}
\end{equation}
\begin{equation}
{L} =  {L}_{rgb} + \lambda_{7} {L}_{flatten} + {L}_{mask} + {L}_{geo}.
  \label{eq:L}
\end{equation}

\subsection{Mesh Extraction and Gaussians Merging} \label{Mesh Extraction}
After reconstruction, we adopt a method proposed by Zeng et al. \cite{zeng20173dmatch} to fuse depth maps and images into a projective Truncated Signed Distance Function (TSDF) volume and generate the mesh. To obtain a Gaussian model of the whole scene, we merge each sub-scene by position and delete Gaussian primitives outside the bounding box. Due to the expanded boundary of each cell, there is no noticeable seam.

\begin{figure*}[ht] 
	\centering
        \vspace{3 pt}
	\includegraphics[width=\linewidth,scale=1.00]{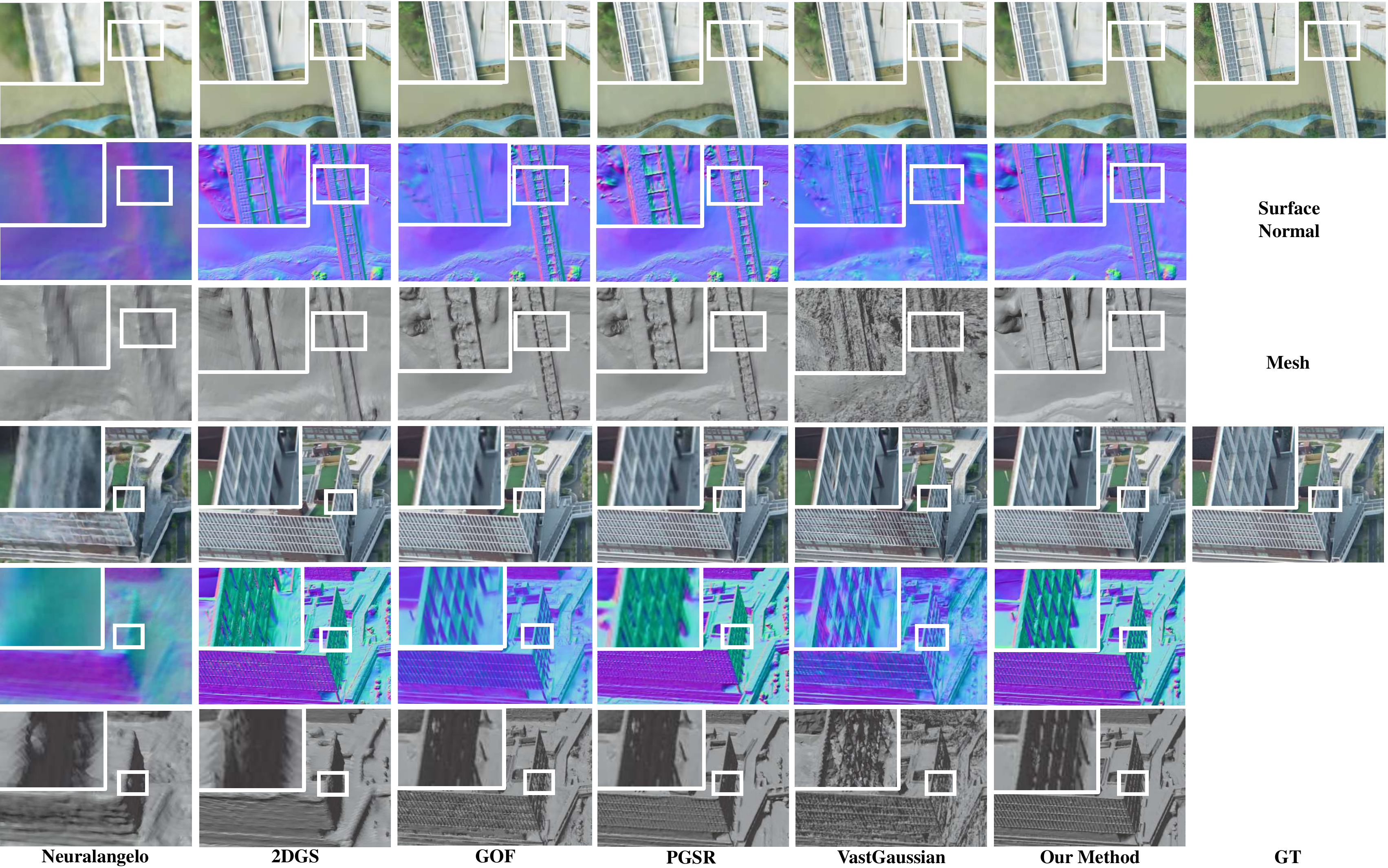}
        \vspace{-20 pt}
	\caption{Qualitative comparison of details between our approach and previous methods. Our results exhibit smooth planes and sharp edges, effectively capturing intricate details from full-size images.}
        \vspace{-15 pt}
	\label{Figure_4}
\end{figure*}

\section{Experiments}

\subsection{Experimental Setup}
\paragraph{Implementation and Dataset} In our experiments, coarse and fine optimization share the 30,000 iterations. Images are downsampled four times in coarse training, facilitating the reconstruction of overall structures. In partitioning and refining, images are split into four sub-images, reducing the computational consumption while keeping details. The learning rates of appearance embedding, CNN, and the Unet model are all 0.001. We comprehensively evaluated seven scenes from a publicly available dataset, GauU-Scene V2 \cite{xiong2024gauu}. These scenes range from 0.8 to 1.8 $\text{km}^2$, and images are captured by Unmanned Aerial Vehicles (UAV). All scenes have LiDAR point clouds, which could be used to assess the accuracy of surface reconstruction. The camera poses and sparse point cloud were obtained by colmap \cite{schonberger2016structure}. Based on colmap results, the maximum image number threshold is 500, and the minimum length threshold is 3 for partitioning. We set $\lambda_{7} = 100$ to enforce the Gaussians being flattened. For geometric loss, we set $\lambda_{3} = 0.01$, $\lambda_{4} = \lambda_{6} = 0.2$, and $\lambda_{5} = 0.05$ in all experiments. We conducted the ablation research on the Russian Building.

\paragraph{Baseline and Metrics} We compared the our visual and geometric results with Neuralangelo \cite{li2023neuralangelo}, PGSR \cite{chen2024pgsr}, GOF \cite{yu2024gaussian}, 2DGS \cite{huang20242d} and VastGaussian \cite{lin2024vastgaussian}. We maintained the default parameters for PGSR, GOF, 2DGS, and VastGaussian and downsampled images four times for Neuralangelo to avoid being out of memory. All experiments were conducted on a server with A6000 GPUs. To assess the performance in detail, we utilized full-resolution rendered images to calculate the PSNR, SSIM \cite{wang2004image}, and LPIPS \cite{zhang2018unreasonable}. Mean Absolute Error (MAE) and Root Mean Square Error (RMSE) between the generated mesh and LiDAR point clouds are employed to evaluate the geometric accuracy.

\subsection{Results}

\begin{table}[t]
\centering
\caption{Ablation on partitioning strategy.}
\vspace{-5 pt}
\begin{threeparttable}
\scalebox{0.95}{\begin{tabular}{l|ccccc}
\toprule[1pt]
                & \#Cell      & PSNR$\uparrow$ & MAE$\downarrow$ & Average Time(h)$\downarrow$ \\
\midrule
w/o Size Thre\tnote{a} & 34          & 24.58          & \textbf{0.63}          & \textbf{9.75} \\
w/o Num Thre\tnote{b} & 28        & 25.26          & 0.67          & 10.21         \\
Regular Partitioning & 16        & 25.48          & 0.70          & 10.97         \\
\midrule
Full Model      & 18        & \textbf{25.79} & 0.67          & 10.60         \\
\bottomrule[1pt]
\end{tabular}}
\begin{tablenotes}
        \footnotesize
        \item[a] without minimum size threshold.
        \item[b] without maximum image number threshold.
\end{tablenotes}
\vspace{-20 pt}
\label{Table_3}
\end{threeparttable}
\end{table}

\paragraph{Geometic Reconstruction}  As shown in Fig. \ref{Figure_4}, our method captures the highest level of detail, including clearly regular structures and thin objects, from full-resolution images. While other methods tend to oversmooth the details. The meshing method of 2DGS \cite{huang20242d} fails to extract an accurate mesh from the Gaussian model. GOF \cite{yu2024gaussian} and VastGaussian \cite{lin2024vastgaussian} show noisy surfaces with holes. Tab. \ref{Table_1} indicates that our method outperforms explicit and implicit methods in most scenes and achieves the overall highest accuracy regarding the mean MAE and RMSE. Due to a lack of supervision at the boundary of scenes and insufficient pruning mechanism, the Gaussian primitives grow uncontrolled throughout the space, indicating large geometric errors, especially for the Morden Building scene. While PGSR and our method avoid it thanks to multi-view constraints. 

\begin{figure}[t] 
	\centering
        \vspace{3 pt}
	\includegraphics[width=\linewidth,scale=1.00]{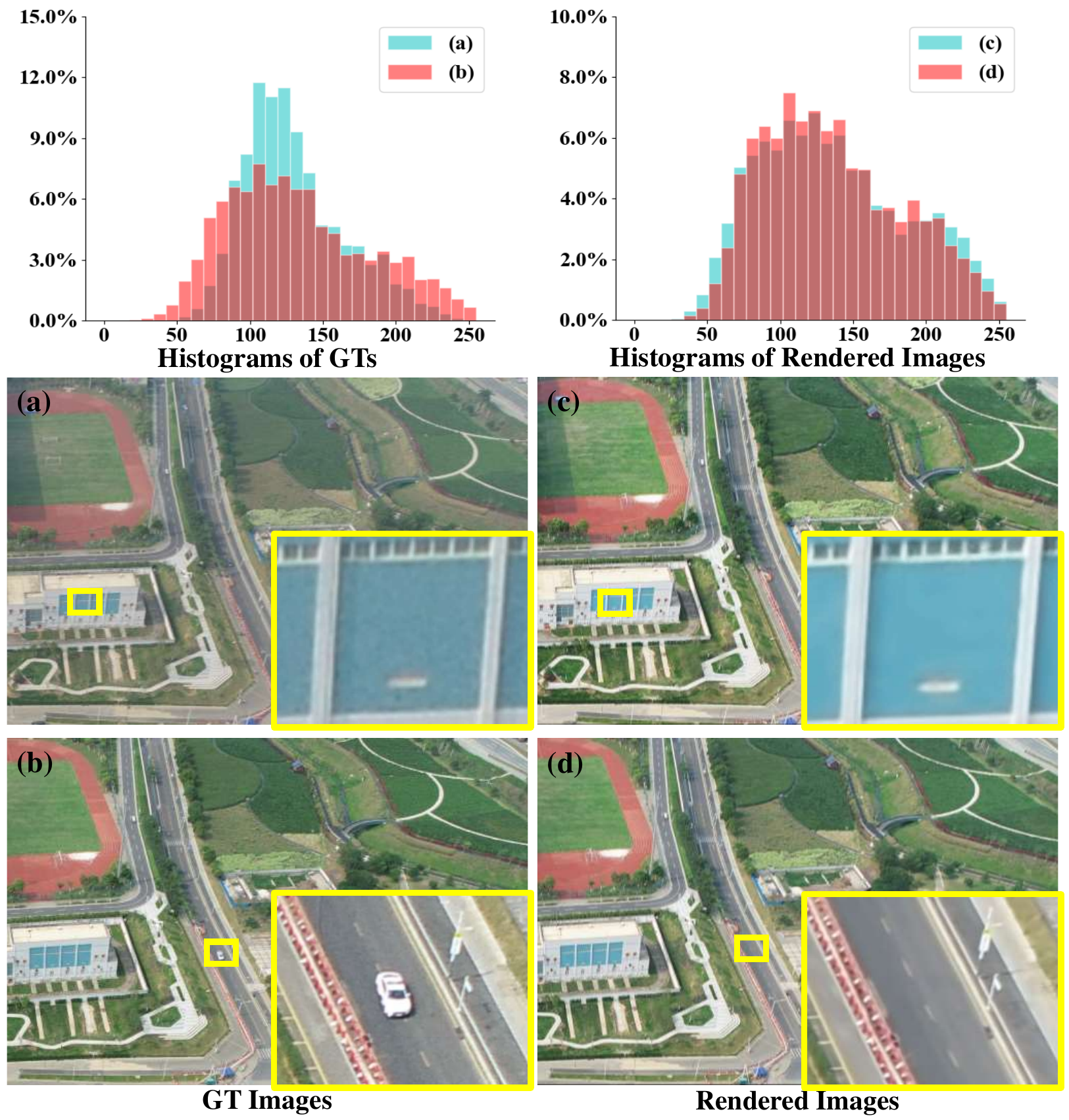}
        \vspace{-20 pt}
	\caption{ Visualization of dynamic appearance decoupling with the corresponding histograms.}
         \vspace{-10 pt}
	\label{Figure_5}
\end{figure}

\paragraph{Visual Quality} Fig. \ref{Figure_4} demonstrates that our method obtains visually pleasing results with thin structure. Other methods exhibit blurry results or lack of details. As shown in the histograms of Fig. \ref{Figure_5}, the rendered images have a globally consistent appearance compared to GTs. The second row of Fig. \ref{Figure_5} shows a cleaner rendered image compared with the noisy GT, and the transient mask model successfully identifies and removes the moving cars in the last row. We report SSIM, PSNR, and LPIPS in Tab. \ref{Table_2}, showing that our method achieves the most photo-realistic results in most cases and the overall best performance. 

\begin{figure}[t] 
	\centering
	\includegraphics[width=\linewidth,scale=1.00]{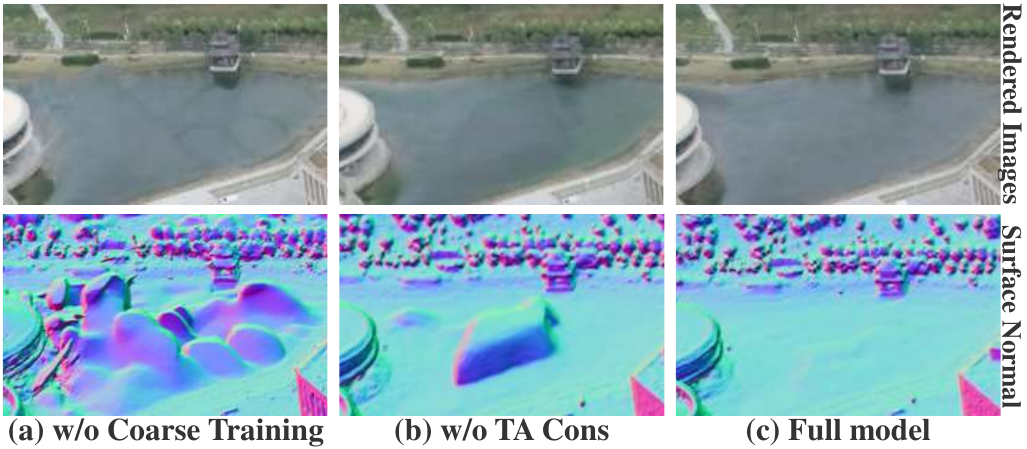}
        \vspace{-20 pt}
	\caption{ Ablation on coarse training (a) and texture-less area constraint (b). }
         \vspace{-15 pt}
	\label{Figure_6}
\end{figure}

\subsection{Ablation}

\paragraph{Partitioning Strategy} We report the number of cells, PSNR, MAE, and average time of reconstructing one cell in Tab. \ref{Table_3} with different partitioning strategies. With more cells, the geometric accuracy tends to be higher while the visual quality decreases, with a drop in consumption time. While with a similar number of cells, our adaptive method can reasonably assign images and reconstruction range to each cell, achieving better rendering and geometric results.

\paragraph{Dynamic Appearance Decoupling} Table \ref{Table_4} presents the MAE, RMSE, and the number of Gaussian primitives to evaluate the dynamic appearance module. This module significantly reduces the number of Gaussians, saving GPU memory. The absence of this module leads to floaters at scene boundaries due to insufficient appearance consistency and limited views, which subsequently amplify geometric errors.

\paragraph{Geometry Regularization} As shown in Tab. \ref{Table_5}, the multi-view constraint effectively ensures geometric consistency while slightly decreasing the visual quality. The texture-less area constraint smooths the water area, as displayed in Fig. \ref{Figure_6}. This kind of error is hardly observed in the second row of Tab. \ref{Table_5} because the LiDAR device can not capture the surface of water or glass. Noticing the relatively low resolution of the point cloud and the systematic error between the point cloud and the actual surface, our improvement may be obscured.

\paragraph{Coarse-to-fine Strategy} As shown in Tab. \ref{Table_5}, coarse training provides an overall geometric structure and prevents getting stuck in a locally optimal solution. Fine training improves the detail with higher PSNR and geometric accuracy. Fig. \ref{Figure_6}(a) demonstrates that the texture-less areas become noisy with floaters without coarse training. 

\begin{table}[t]
\centering
\vspace{4 pt}
\caption{Ablation on dynamic appearance decoupling model.}
\vspace{-10 pt}
\begin{threeparttable}
\scalebox{0.95}{\begin{tabular}{l|ccc}
\toprule[1pt]
                  & MAE$\downarrow$ & RMSE$\downarrow$ & \#Gaussians(million)$\downarrow$ \\
\midrule
w/o Decoupling AM \tnote{a} & 1.05    & 3.86   & 13.74         \\
w/o Transient MM \tnote{b} & 1.15    & 3.93   & 19.51         \\
\midrule
Full Model        & \textbf{0.67}    & \textbf{1.43}   & \textbf{12.25}         \\
\bottomrule[1pt]
\end{tabular}}
\begin{tablenotes}
        \footnotesize
        \item[a] without decoupling appearance model.
        \item[b] without transient mask model.
\end{tablenotes}
\vspace{-10 pt}
\label{Table_4}
\end{threeparttable}
\end{table}

\begin{table}[t]
\centering
\caption{Ablation on geometry regularization and fine/coarse training.}
\vspace{-10 pt}
\begin{threeparttable}
\scalebox{0.95}{\begin{tabular}{l|ccccc}
\toprule[1pt]
                   & SSIM $\uparrow$    & PSNR$\uparrow$  & LPIPS$\downarrow$   & MAE$\downarrow$  & RMSE$\downarrow$      \\
\midrule
w/o MV Cons \tnote{a} & \textbf{0.70} & \textbf{25.90} & \textbf{0.40} & 1.21         & 3.68             \\
w/o TA Cons \tnote{b} & 0.66          & 24.72          & 0.43          & \textbf{0.64}   & \textbf{1.33}   \\
w/o Fine  & 0.63           & 23.31          & 0.47           & 0.69            & 1.44            \\
w/o Coarse  & 0.65           & 23.92         & 0.44           & 0.79    & 1.49             \\
\midrule
Full Model  & \underline{0.68}   & \underline{25.79}  & \underline{0.41}     & \underline{0.67} & \underline{1.43} \\
\bottomrule[1pt]
\end{tabular}}
\begin{tablenotes}
        \footnotesize
        \item[a] without multi-view constraint.
        \item[b] without texture-less area constraint.
\end{tablenotes}
\vspace{-15 pt}
\label{Table_5}
\end{threeparttable}
\end{table}
\section{Conclusion and Limitations }
\label{sec:conclusion}

In this paper, we propose a novel, fine-detailed surface reconstruction method for large-scale scenes from full-size images. The proposed adaptive partitioning strategy based on image number and cell size can evenly distribute computational resources to sub-scenes. Our dynamic appearance model decouples the global appearance variance and eliminates the influence from transient objects, enabling consistent rendering and geometry. The constraints directly to geometry and optimization from full-sized images ensure accurate surfaces with fine details. The experiments on the GauU-scene V2 dataset show that our method achieves accurate geometric surface and the highest visual quality, exceeding the performance of NeRF-based and Gaussian-based methods. Although our method achieves outstanding results, we do not consider reflective or transparent surfaces, such as glass, which are common in urban areas. The fine-detailed surfaces are fitted by a large number of Gaussian primitives and supervised by high-resolution images, requiring a large amount of reconstruction time.








\bibliographystyle{IEEEtran}
\bibliography{References}

\begin{thebibliography}{10}
\providecommand{\url}[1]{#1}
\csname url@rmstyle\endcsname
\providecommand{\newblock}{\relax}
\providecommand{\bibinfo}[2]{#2}
\providecommand\BIBentrySTDinterwordspacing{\spaceskip=0pt\relax}
\providecommand\BIBentryALTinterwordstretchfactor{4}
\providecommand\BIBentryALTinterwordspacing{\spaceskip=\fontdimen2\font plus
\BIBentryALTinterwordstretchfactor\fontdimen3\font minus
  \fontdimen4\font\relax}
\providecommand\BIBforeignlanguage[2]{{%
\expandafter\ifx\csname l@#1\endcsname\relax
\typeout{** WARNING: IEEEtran.bst: No hyphenation pattern has been}%
\typeout{** loaded for the language `#1'. Using the pattern for}%
\typeout{** the default language instead.}%
\else
\language=\csname l@#1\endcsname
\fi
#2}}

\bibitem{Vizzo2021Poisson}
I.~Vizzo, X.~Chen, N.~Chebrolu, J.~Behley, and C.~Stachniss, ``Poisson surface
  reconstruction for lidar odometry and mapping,'' in \emph{2021 IEEE
  International Conference on Robotics and Automation (ICRA)}, 2021, pp.
  5624--5630.

\bibitem{Liu2019HighDefinitionMap}
R.~Liu, J.~Wang, and B.~Zhang, ``High definition map for automated driving:
  Overview and analysis,'' \emph{Journal of Navigation}, vol.~73, pp. 1--18, 08
  2019.

\bibitem{tancik2022block}
M.~Tancik, V.~Casser, X.~Yan, S.~Pradhan, B.~Mildenhall, P.~P. Srinivasan,
  J.~T. Barron, and H.~Kretzschmar, ``Block-nerf: Scalable large scene neural
  view synthesis,'' in \emph{Proceedings of the IEEE/CVF Conference on Computer
  Vision and Pattern Recognition}, 2022, pp. 8248--8258.

\bibitem{Jun2024RenderableStreetView}
H.~Jun, H.~Yu, and S.~Oh, ``Renderable street view map-based localization:
  Leveraging 3d gaussian splatting for street-level positioning,'' in
  \emph{2024 IEEE/RSJ International Conference on Intelligent Robots and
  Systems (IROS)}, 2024, pp. 5635--5640.

\bibitem{Yan2024StreetGaussians}
Y.~Yan, H.~Lin, C.~Zhou, W.~Wang, H.~Sun, K.~Zhan, X.~Lang, X.~Zhou, and
  S.~Peng, ``Street gaussians: Modeling dynamic urban scenes with gaussian
  splatting,'' in \emph{Computer Vision -- ECCV 2024}, A.~Leonardis, E.~Ricci,
  S.~Roth, O.~Russakovsky, T.~Sattler, and G.~Varol, Eds.\hskip 1em plus 0.5em
  minus 0.4em\relax Cham: Springer Nature Switzerland, 2025, pp. 156--173.

\bibitem{chen2019learning}
Z.~Chen and H.~Zhang, ``Learning implicit fields for generative shape
  modeling,'' in \emph{Proceedings of the IEEE/CVF conference on computer
  vision and pattern recognition}, 2019, pp. 5939--5948.

\bibitem{mildenhall2021nerf}
B.~Mildenhall, P.~P. Srinivasan, M.~Tancik, J.~T. Barron, R.~Ramamoorthi, and
  R.~Ng, ``Nerf: Representing scenes as neural radiance fields for view
  synthesis,'' \emph{Communications of the ACM}, vol.~65, no.~1, pp. 99--106,
  2021.

\bibitem{barron2021mip}
J.~T. Barron, B.~Mildenhall, M.~Tancik, P.~Hedman, R.~Martin-Brualla, and P.~P.
  Srinivasan, ``Mip-nerf: A multiscale representation for anti-aliasing neural
  radiance fields,'' in \emph{Proceedings of the IEEE/CVF international
  conference on computer vision}, 2021, pp. 5855--5864.

\bibitem{muller2022instant}
T.~M{\"u}ller, A.~Evans, C.~Schied, and A.~Keller, ``Instant neural graphics
  primitives with a multiresolution hash encoding,'' \emph{ACM transactions on
  graphics (TOG)}, vol.~41, no.~4, pp. 1--15, 2022.

\bibitem{turki2022mega}
H.~Turki, D.~Ramanan, and M.~Satyanarayanan, ``Mega-nerf: Scalable construction
  of large-scale nerfs for virtual fly-throughs,'' in \emph{Proceedings of the
  IEEE/CVF Conference on Computer Vision and Pattern Recognition}, 2022, pp.
  12\,922--12\,931.

\bibitem{NEURIPS2021_e41e164f}
P.~Wang, L.~Liu, Y.~Liu, C.~Theobalt, T.~Komura, and W.~Wang, ``Neus: Learning
  neural implicit surfaces by volume rendering for multi-view reconstruction,''
  in \emph{Advances in Neural Information Processing Systems}, M.~Ranzato,
  A.~Beygelzimer, Y.~Dauphin, P.~Liang, and J.~W. Vaughan, Eds., vol.~34.\hskip
  1em plus 0.5em minus 0.4em\relax Curran Associates, Inc., 2021, pp.
  27\,171--27\,183.

\bibitem{fu2022geo}
Q.~Fu, Q.~Xu, Y.~S. Ong, and W.~Tao, ``Geo-neus: Geometry-consistent neural
  implicit surfaces learning for multi-view reconstruction,'' \emph{Advances in
  Neural Information Processing Systems}, vol.~35, pp. 3403--3416, 2022.

\bibitem{li2023neuralangelo}
Z.~Li, T.~M{\"u}ller, A.~Evans, R.~H. Taylor, M.~Unberath, M.-Y. Liu, and C.-H.
  Lin, ``Neuralangelo: High-fidelity neural surface reconstruction,'' in
  \emph{Proceedings of the IEEE/CVF Conference on Computer Vision and Pattern
  Recognition}, 2023, pp. 8456--8465.

\bibitem{kerbl20233d}
B.~Kerbl, G.~Kopanas, T.~Leimk{\"u}hler, and G.~Drettakis, ``3d gaussian
  splatting for real-time radiance field rendering.'' \emph{ACM Trans. Graph.},
  vol.~42, no.~4, pp. 139--1, 2023.

\bibitem{lin2024gaussian}
Y.~Lin, Z.~Dai, S.~Zhu, and Y.~Yao, ``Gaussian-flow: 4d reconstruction with
  dynamic 3d gaussian particle,'' in \emph{Proceedings of the IEEE/CVF
  Conference on Computer Vision and Pattern Recognition}, 2024, pp.
  21\,136--21\,145.

\bibitem{yu2024mip}
Z.~Yu, A.~Chen, B.~Huang, T.~Sattler, and A.~Geiger, ``Mip-splatting:
  Alias-free 3d gaussian splatting,'' in \emph{Proceedings of the IEEE/CVF
  Conference on Computer Vision and Pattern Recognition}, 2024, pp.
  19\,447--19\,456.

\bibitem{guedon2024sugar}
A.~Gu{\'e}don and V.~Lepetit, ``Sugar: Surface-aligned gaussian splatting for
  efficient 3d mesh reconstruction and high-quality mesh rendering,'' in
  \emph{Proceedings of the IEEE/CVF Conference on Computer Vision and Pattern
  Recognition}, 2024, pp. 5354--5363.

\bibitem{huang20242d}
B.~Huang, Z.~Yu, A.~Chen, A.~Geiger, and S.~Gao, ``2d gaussian splatting for
  geometrically accurate radiance fields,'' in \emph{ACM SIGGRAPH 2024
  Conference Papers}, 2024, pp. 1--11.

\bibitem{chen2024pgsr}
D.~Chen, H.~Li, W.~Ye, Y.~Wang, W.~Xie, S.~Zhai, N.~Wang, H.~Liu, H.~Bao, and
  G.~Zhang, ``Pgsr: Planar-based gaussian splatting for efficient and
  high-fidelity surface reconstruction,'' \emph{IEEE Transactions on
  Visualization and Computer Graphics}, pp. 1--12, 2024.

\bibitem{yu2024gaussian}
Z.~Yu, T.~Sattler, and A.~Geiger, ``Gaussian opacity fields: Efficient adaptive
  surface reconstruction in unbounded scenes,'' \emph{ACM Transactions on
  Graphics (TOG)}, vol.~43, no.~6, pp. 1--13, 2024.

\bibitem{lin2024vastgaussian}
J.~Lin, Z.~Li, X.~Tang, J.~Liu, S.~Liu, J.~Liu, Y.~Lu, X.~Wu, S.~Xu, Y.~Yan,
  \emph{et~al.}, ``Vastgaussian: Vast 3d gaussians for large scene
  reconstruction,'' in \emph{Proceedings of the IEEE/CVF Conference on Computer
  Vision and Pattern Recognition}, 2024, pp. 5166--5175.

\bibitem{liu2024citygaussian}
Y.~Liu, C.~Luo, L.~Fan, N.~Wang, J.~Peng, and Z.~Zhang, ``Citygaussian:
  Real-time high-quality large-scale scene rendering with gaussians,'' in
  \emph{Computer Vision -- ECCV 2024}, A.~Leonardis, E.~Ricci, S.~Roth,
  O.~Russakovsky, T.~Sattler, and G.~Varol, Eds.\hskip 1em plus 0.5em minus
  0.4em\relax Cham: Springer Nature Switzerland, 2025, pp. 265--282.

\bibitem{chen2024dogaussian}
Y.~Chen and G.~H. Lee, ``Dogs: Distributed-oriented gaussian splatting for
  large-scale 3d reconstruction via gaussian consensus,'' in \emph{The
  Thirty-eighth Annual Conference on Neural Information Processing Systems},
  2024.

\bibitem{ronneberger2015u}
O.~Ronneberger, P.~Fischer, and T.~Brox, ``U-net: Convolutional networks for
  biomedical image segmentation,'' in \emph{Medical image computing and
  computer-assisted intervention--MICCAI 2015: 18th international conference,
  Munich, Germany, October 5-9, 2015, proceedings, part III 18}.\hskip 1em plus
  0.5em minus 0.4em\relax Springer, 2015, pp. 234--241.

\bibitem{schonberger2016pixelwise}
J.~L. Sch{\"o}nberger, E.~Zheng, J.-M. Frahm, and M.~Pollefeys, ``Pixelwise
  view selection for unstructured multi-view stereo,'' in \emph{Computer Vision
  -- ECCV 2016}, B.~Leibe, J.~Matas, N.~Sebe, and M.~Welling, Eds.\hskip 1em
  plus 0.5em minus 0.4em\relax Cham: Springer International Publishing, 2016,
  pp. 501--518.

\bibitem{fridovich2022plenoxels}
S.~Fridovich-Keil, A.~Yu, M.~Tancik, Q.~Chen, B.~Recht, and A.~Kanazawa,
  ``Plenoxels: Radiance fields without neural networks,'' in \emph{Proceedings
  of the IEEE/CVF conference on computer vision and pattern recognition}, 2022,
  pp. 5501--5510.

\bibitem{wu20244d}
G.~Wu, T.~Yi, J.~Fang, L.~Xie, X.~Zhang, W.~Wei, W.~Liu, Q.~Tian, and X.~Wang,
  ``4d gaussian splatting for real-time dynamic scene rendering,'' in
  \emph{Proceedings of the IEEE/CVF Conference on Computer Vision and Pattern
  Recognition}, 2024, pp. 20\,310--20\,320.

\bibitem{fan2023lightgaussian}
Z.~Fan, K.~Wang, K.~Wen, Z.~Zhu, D.~Xu, and Z.~Wang, ``Lightgaussian: Unbounded
  3d gaussian compression with 15x reduction and 200+ fps,'' \emph{arXiv
  preprint arXiv:2311.17245}, 2023.

\bibitem{oechsle2021unisurf}
M.~Oechsle, S.~Peng, and A.~Geiger, ``Unisurf: Unifying neural implicit
  surfaces and radiance fields for multi-view reconstruction,'' in
  \emph{Proceedings of the IEEE/CVF International Conference on Computer
  Vision}, 2021, pp. 5589--5599.

\bibitem{zhenxing2022switch}
M.~Zhenxing and D.~Xu, ``Switch-nerf: Learning scene decomposition with mixture
  of experts for large-scale neural radiance fields,'' in \emph{The Eleventh
  International Conference on Learning Representations}, 2022.

\bibitem{xu2023grid}
L.~Xu, Y.~Xiangli, S.~Peng, X.~Pan, N.~Zhao, C.~Theobalt, B.~Dai, and D.~Lin,
  ``Grid-guided neural radiance fields for large urban scenes,'' in
  \emph{Proceedings of the IEEE/CVF Conference on Computer Vision and Pattern
  Recognition}, 2023, pp. 8296--8306.

\bibitem{xiangli2022bungeenerf}
Y.~Xiangli, L.~Xu, X.~Pan, N.~Zhao, A.~Rao, C.~Theobalt, B.~Dai, and D.~Lin,
  ``Bungeenerf: Progressive neural radiance field for extreme multi-scale scene
  rendering,'' in \emph{Computer Vision -- ECCV 2022}, S.~Avidan, G.~Brostow,
  M.~Ciss{\'e}, G.~M. Farinella, and T.~Hassner, Eds.\hskip 1em plus 0.5em
  minus 0.4em\relax Cham: Springer Nature Switzerland, 2022, pp. 106--122.

\bibitem{liu2024citygaussianv2}
\BIBentryALTinterwordspacing
Y.~Liu, C.~Luo, Z.~Mao, J.~Peng, and Z.~Zhang, ``Citygaussianv2: Efficient and
  geometrically accurate reconstruction for large-scale scenes,'' 2024.
  [Online]. Available: \url{https://arxiv.org/abs/2411.00771}
\BIBentrySTDinterwordspacing

\bibitem{Zhang2024Gaussianinthewild}
D.~Zhang, C.~Wang, W.~Wang, P.~Li, M.~Qin, and H.~Wang, ``Gaussian in the wild:
  3d gaussian splatting for unconstrained image collections,'' in
  \emph{Computer Vision -- ECCV 2024}, A.~Leonardis, E.~Ricci, S.~Roth,
  O.~Russakovsky, T.~Sattler, and G.~Varol, Eds.\hskip 1em plus 0.5em minus
  0.4em\relax Cham: Springer Nature Switzerland, 2025, pp. 341--359.

\bibitem{yoo2009fast}
J.-C. Yoo and T.~H. Han, ``Fast normalized cross-correlation,'' \emph{Circuits,
  systems and signal processing}, vol.~28, pp. 819--843, 2009.

\bibitem{zeng20173dmatch}
A.~Zeng, S.~Song, M.~Nie{\ss}ner, M.~Fisher, J.~Xiao, and T.~Funkhouser,
  ``3dmatch: Learning local geometric descriptors from rgb-d reconstructions,''
  in \emph{Proceedings of the IEEE conference on computer vision and pattern
  recognition}, 2017, pp. 1802--1811.

\bibitem{xiong2024gauu}
B.~Xiong, N.~Zheng, J.~Liu, and Z.~Li, ``Gauu-scene v2: Assessing the
  reliability of image-based metrics with expansive lidar image dataset using
  3dgs and nerf,'' \emph{CoRR}, 2024.

\bibitem{schonberger2016structure}
J.~L. Schonberger and J.-M. Frahm, ``Structure-from-motion revisited,'' in
  \emph{Proceedings of the IEEE conference on computer vision and pattern
  recognition}, 2016, pp. 4104--4113.

\bibitem{wang2004image}
Z.~Wang, A.~C. Bovik, H.~R. Sheikh, and E.~P. Simoncelli, ``Image quality
  assessment: from error visibility to structural similarity,'' \emph{IEEE
  transactions on image processing}, vol.~13, no.~4, pp. 600--612, 2004.

\bibitem{zhang2018unreasonable}
R.~Zhang, P.~Isola, A.~A. Efros, E.~Shechtman, and O.~Wang, ``The unreasonable
  effectiveness of deep features as a perceptual metric,'' in \emph{Proceedings
  of the IEEE conference on computer vision and pattern recognition}, 2018, pp.
  586--595.

\end{thebibliography}


\end{document}